\documentclass[conference]{IEEEtran}
\IEEEoverridecommandlockouts
\usepackage{amsmath,amssymb,amsfonts}
\usepackage{algorithmic}
\usepackage{graphicx}
\usepackage{textcomp}
\usepackage{xcolor}
\usepackage{balance}
\usepackage{multirow}
\usepackage[square,sort,comma,numbers]{natbib}
\usepackage{outlines}
\usepackage{physics}
\usepackage{tikz}
\usetikzlibrary{quantikz}
\usepackage{braket}
\usepackage[bf]{caption} 
\usepackage{adjustbox}
\usepackage{subcaption}

\usepackage{orcidlink}
\def\BibTeX{{\rm B\kern-.05em{\sc i\kern-.025em b}\kern-.08em
    T\kern-.1667em\lower.7ex\hbox{E}\kern-.125emX}}
\begin{document}

\title{An Efficient Quantum Binary-Neuron Algorithm for Accurate Multi-Story Floor Localization}

\author{
\IEEEauthorblockN{Yousef Zook}
\IEEEauthorblockA{\textit{Computer and Systems Engineering}}
\textit{Alexandria University}\\
Alexandria, Egypt \\
es-yousif.mohamed@alexu.edu.eg\orcidlink{0000-0003-1105-5464}
\and
\IEEEauthorblockN{Ahmed Shokry}
\IEEEauthorblockA{\textit{Computer Science and Engineering}}
\textit{Pennsylvania State University}\\
PA, USA \\
ahmed.shokry@psu.edu\orcidlink{0000-0003-3753-8886}
\and
\IEEEauthorblockN{Moustafa Youssef}
\IEEEauthorblockA{\textit{Computer Science and Engineering}}
\textit{University of New South Wales} \\
Sydney, Australia \\
m.yousief@unsw.edu.au\orcidlink{0000-0002-2063-4364}
}

\maketitle

\begin{abstract}
 Accurate floor localization in a multi-story environment is an important but challenging task. Among the current floor localization techniques, fingerprinting is the mainstream technology due to its accuracy in noisy environments. To achieve accurate floor localization in a building with many floors, we have to collect sufficient data on each floor, which needs significant storage and running time; preventing fingerprinting techniques from scaling to support large multi-story buildings, especially on a worldwide scale. %This becomes even more crucial for many  emergency-based applications that need a fast accurate response.

In this paper, we propose a quantum algorithm for accurate multi-story localization. The proposed algorithm leverages  quantum computing concepts to provide an exponential enhancement in both space and running time compared to the classical counterparts. In addition, it builds on an efficient binary-neuron implementation that can be implemented using fewer qubits compared to the typical non-binary neurons, allowing for easier deployment with  near-term quantum devices. %xxx should we put near-term quantum in the title?
 We implement the proposed algorithm on a real IBM quantum machine and evaluate it on three real indoor testbeds. Results confirm the exponential saving in both time and space for the proposed quantum algorithm, while keeping the same localization accuracy compared to the traditional classical techniques, and using half the number of qubits required for other  quantum localization algorithms. % leading to better accuracy on near-term quantum devices.
\end{abstract}


\section{Introduction}

 Indoor localization  has become an important requirement in smart cities and Internet of Things (IoT)-based applications as people spend more time indoors~\cite{wan2022self}. Accurate 3D localization is crucial in many commercial and emergency situations, such as indoor multi-floor navigation and emergency response services, where getting an accurate location in a large multi-story building can make the difference between life and death~\cite{saeed2022cellstory}. 

Different techniques have been proposed to solve the floor estimation problem~\cite{main_classical_paper, fi_fingerprint_2018, fi_sens_acm_18_my, survey_multi_floor}. 
 Among the current floor localization techniques, fingerprinting is the mainstream technology due to its accuracy in noisy environments~\cite{main_classical_paper}.  These techniques have two phases: the offline calibration phase and the online phase. In the offline phase, the data coming from the different signal sources (e.g. WiFi access points, APs) is collected at the different floors in the building. Then, in the online phase, the system matches the heard signals from a user's device  at an unknown floor with the fingerprint data to estimate her floor. The closest floor in the fingerprint is reported as the estimated floor.

 The number of signal sources used in the matching process affects the final floor estimation accuracy; i.e., the higher the number of signal sources, the more precise the accuracy will be\cite{bahl2000radar, youssef2005horus}. However, this increases the matching time and the space needed to store a large fingerprint, especially for large multi-story buildings.
This drawback poses a major difficulty for fingerprinting techniques in real-world floor localization systems.

 Recently, quantum fingerprint-based localization algorithms started to appear to enable a large-scale worldwide localization~\cite{quantum_arx, quantum_qce, quantum_lcn, quantum_vision, logMN_paper, eqes, deployable}, mainly focusing on \textbf{\textit{2D localization in a single floor}}. However, these algorithms require a relatively large number of qubits, making them less suitable for near-term quantum devices. This is even \textit{\textbf{worse}} when we apply them to the multi-floor localization problem, where the \textit{fingerprint is much larger}.

 In this paper, we propose an efficient quantum fingerprint-based floor localization algorithm for near-term quantum devices. The proposed  algorithm uses the quantum binary neuron to provide an exponential saving in time and space compared to its classical counterparts~\cite{quantum_neuron_paper}. Furthermore, this implementation requires nearly \textit{half} the number of qubits compared to current quantum localization algorithms~\cite{quantum_arx, quantum_qce, quantum_lcn, quantum_vision}. This allows our algorithm to use double the number of signal sources, leading to better localization accuracy and/or more efficient implementation. 

 We implement our quantum algorithm on an IBM Quantum Experience machine and deploy it in three real multi-story buildings. The results confirm the ability of our quantum algorithm to achieve the same accuracy as classical floor localization but with the potential exponential saving in both space and running time. 

The rest of the paper is organized as follows: Section~\ref{sec:background}  gives a background on the floor localization problem and the binary neuron. We discuss the details of the proposed quantum floor localization algorithm in Section~\ref{sec:algorithm}. Then, we evaluate the performance of our system in Section~\ref{sec:evaluation}. Finally, we conclude our work in Section~\ref{sec:conclusion}.
%

%%%%%%%%%%%%%%%%%%%%%%%%%%%%%%%%%%%%%%%%%%%%%%%%%%%%%%%%%%%%%%%%%%%%%%%%%%%%%%%%%%%%%%%%%%%%%%%%%%%%%%%%%%%%%%%%%%%%%
\section{Background}
\label{sec:background}

\subsection{The Floor Estimation Problem}
Fingerprint-based floor localization algorithms~\cite{main_classical_paper} depend on the pre-installed signal sources (e.g., WiFi or cellular infrastructure, also called reference points) to estimate users' floor. Generally, they work in two phases: the offline calibration phase, where the signal information, such as the Received Signal Strength (RSS) and the signal source ID, is collected at different locations on each floor to construct the fingerprint database. The fingerprint is stored as vectors, where each entry in the vector represents the RSS from one reference point, alongside the floor number where these vectors are collected. In the online tracking phase, the online signal sources information is matched to the fingerprint vectors, and the closest floor in the signal space becomes the estimated floor.  
\textit{Classically}, matching the online information to the fingerprint takes \textbf{\textit{$o(MN)$ time and space complexity}}, where $M$ is the number of vectors in the fingerprint and $N$ is the number of signal sources. 

 As the floor localization accuracy depends on the density of the signal sources, the matching time and space requirements limit the accuracy of the fingerprint-based techniques, especially for large buildings.

\subsection{Binary Neuron}
\begin{figure}[!t]
	\centerline
	{\includegraphics[width=0.3\textwidth]{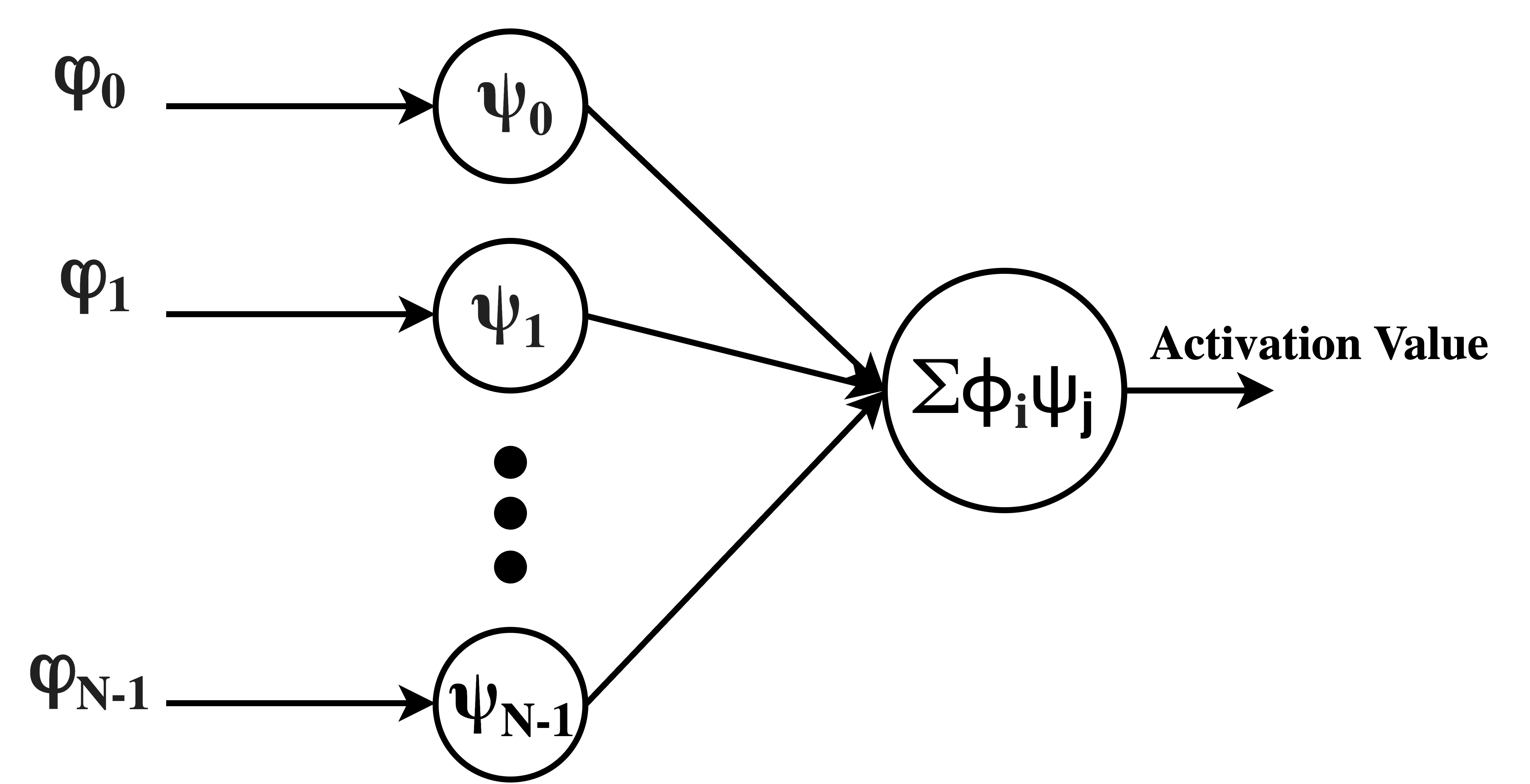}}
	\caption{A general artificial neuron model diagram. The neuron input $\Vec{\Phi}$ is processed with the neuron weight vector $\Vec{\Psi}$ using dot product to produce the output of the neuron (the activation value). A binary neuron has binary inputs and weight vectors.}
	\label{fig:neuron}
\end{figure}

An artificial neuron mimics what is done in a real biological neuron. It takes signals as inputs, processes these signals, and activates an output signal based on this process. An artificial neuron is basically a mathematical representation of the biological neuron (Figure~\ref{fig:neuron}). Specifically, the output activation of a neuron is calculated as the weighted sum of multiplying each input $\phi_i$ with a weight $\psi_i$  as\footnote{The Dirac notation is usually used to mathematically describe quantum systems. The $\ket{.}$ notation
is called ket and represents a column vector, the $\bra{.}$ notation is
called bra and represents a row vector, and the $\braket{.|.}$ notation is
called braket and represents the dot product of the two vectors.}:

\begin{equation}
\label{eq:neuron_dotproduct}
    \braket{\Psi|\Phi} = \sum_{i=0}^{N-1} \Psi_i\Phi_i
\end{equation}

A binary neuron is a special version of the general neuron, where the neuron takes a \textit{binary-valued input} vector $\Vec{\Phi}$, then it is combined with a \textit{binary-valued weight vector} $\Vec{\Psi}$. The output is a response function evaluated from the inner product of the two vectors, representing the neuron \textit{activation} value.

In this paper, we use a quantum implementation of this binary neuron to estimate the user floor number, where the fingerprint samples are the inputs to the neuron and the user's sample represents the neuron weight. The output activation value of this neuron is going to be used as an indicator of the similarity between the user sample and each fingerprint sample. %, where the highest activation value represents highest similarity. 

%\section{Methodology}
\section{The Proposed Quantum Floor Estimation Algorithm}
\label{sec:algorithm}
 To reduce the matching time and space requirements, we use the quantum binary-neuron circuit~\cite{quantum_neuron_paper}.
 The basic idea is to use an instance of the quantum binary neuron to calculate the similarity between the online RSS vector (i.e., user sample) and  each one of the fingerprint training vectors (i.e., fingerprint sample), one at a time. The floor that gives the highest activation is reported as the estimated location for the user sample.

\begin{figure}[!t]
	\centerline
	{\includegraphics[width=0.4\textwidth]{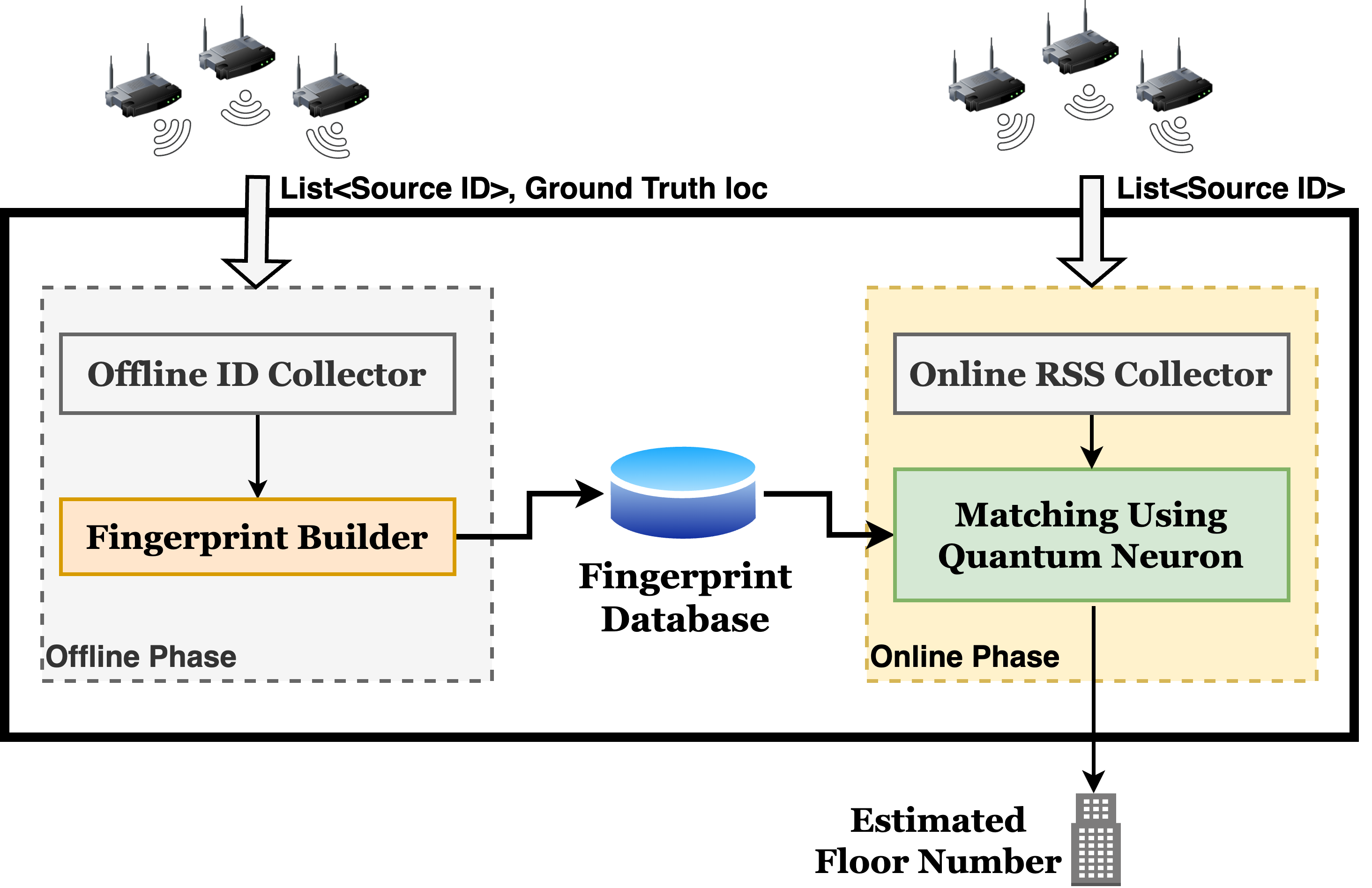}}
	\caption{System architecture for the quantum fingerprint-based floor localization using quantum neuron.}
	\label{fig:arch}
\end{figure}

\subsection{Floor Estimation Algorithm}
 Figure~\ref{fig:arch} shows the architecture of the proposed quantum floor localization system. In the offline phase, the fingerprint data is collected across the building floors by collecting the signal sources IDs along with the floor number as the ground truth in a vector. Each entry in this vector represents a specific signal source. Then, in the online phase, we collect the user signal sources IDs heard by the user's device at an unknown floor. We compare the online RSS vector to vector at each fingerprint location using the quantum binary neuron. The floor in the fingerprint sample with the highest activation value is returned as the estimated floor number.

The RSS vectors are binary vectors: each entry is either $-1$ or $+1$, where $-1$ at a vector index $i$ means that the signal source number $i$ is not heard, and $+1$ represents that the signal source is heard. 

 Assuming that the user online sample is 
 $\Psi=\begin{bmatrix}\psi_0 & ... & \psi_{N-1}\end{bmatrix}$ 
 and the fingerprint sample number $l$ is $\Phi^l=\begin{bmatrix}\phi^l_0 & ... & \phi^l_{N-1}\end{bmatrix}$, where $N$ is the number of the signal sources and $\phi^l_i, \psi_i \in \{-1,1\}$, we can encode the vectors value using amplitude encoding ~\cite{q_preparation} to the following quantum states,

\begin{equation}
\label{eq:states_psi}
    \ket{\Psi} = \frac{1}{\sqrt{N}}\sum_{i=0}^{N-1}\psi_i\ket{i}
\end{equation}

\begin{equation}
\label{eq:states_i}
 \ket{\Phi^l} = \frac{1}{\sqrt{N}}\sum_{i=0}^{N-1}\phi^l_i\ket{i}
\end{equation}
Where $\ket{i}$ is the binary encoding for $i$. 
Note the exponential saving in the space where a vector of $N$ entries is encoded in $\log(N)$ qubits.

 To find the fingerprint sample $\Phi^l$ that is most similar to the online sample $\Psi$, we can use a binary neuron, where each fingerprint sample $\Phi^l$ at location $l$ acts as an input to the neuron and the online sample $\Psi$ acts as the neuron weight vector. Hence, we can find the neuron activation value using the dot product $\braket{\Psi|\Phi^l}$.
%

%%%%%%%%%%%%%%%%%%%%%%%%%%%%%%%%%%%%%Circuit%%%%%%%%%%%%%%%%%%%%%%%%%%%%%%
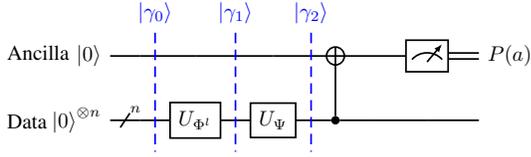
\begin{figure}[!t]
    \centering
    \begin{adjustbox}{width=0.4\textwidth}
    \begin{tikzpicture}[row sep=0.1cm]
			\node[text width=1.5cm] at  (-4,  0.35){Ancilla};
			\node[text width=1.5cm] at  (-4,  -0.8){Data};
                \node[scale=1.0] {
            \begin{quantikz}
                \lstick{$\ket{0}$} & \qw \slice[style={color=blue},label style={color=blue}]{$\ket{\gamma_0}$} &\qw  \slice[style={color=blue},label style={color=blue}]{$\ket{\gamma_1}$} & \qw \slice[style={color=blue},label style={color=blue}]{$\ket{\gamma_2}$} & \targ{}  & \qw &\meter{}& \cw \rstick{$P(a)$}\\
                \lstick{$\ket{0}^{\otimes n}$} & \qwbundle{n} & \gate{U_{\Phi^l}} & \gate{U_{\Psi}} & \ctrl{-1} & \qw & \qw & \qw
            \end{quantikz}};
        \end{tikzpicture}
    \end{adjustbox}
    \caption{Binary neuron quantum circuit to calculate neuron activation between a fingerprint sample $\Phi^l$, and a user sample $\Psi$. $\ket{\gamma_i}$ represents the joint system state at different positions in the circuit.}
    \label{fig:main_circuit}
\end{figure}
%%%%%%%%%%%%%%%%%%%%%%%%%%%%%%%%%%%%%Circuit%%%%%%%%%%%%%%%%%%%%%%%%%%%%%%

 Figure~\ref{fig:main_circuit} shows the quantum circuit for implementing the quantum binary similarity algorithm between the two vectors $\Phi^l$ and $\Psi$ (i.e. $\braket{\Psi|\Phi^l}$). The input to the circuit is a single ancilla qubit at state $\ket{0}$ and $n$-qubit register initially at state $\ket{0}^{\otimes{n}}$, where $N$ is the vectors length representing the $N$ signal sources, and $n = \log(N)$. 

The circuit consists of three stages: $\Phi^l$ Encoding, $\Psi$ Entangling, and Measurement stages. 

\subsubsection{\textbf{The \texorpdfstring{$\Phi^l$}{Lg} Encoding Stage}} Initially, the quantum system is in state:

\begin{equation}
\label{eq:state_0}
    \ket{\gamma_0} = \ket{0}\ket{0}^{\otimes{n}}
\end{equation}
The first step is to construct the fingerprint binary state vector $\Phi^l$. This is achieved by applying a unitary transformation $U_{\Phi^l}$ on the data qubits such that:
\begin{equation}
\label{eq:oracle_i}
    U_{\Phi^l}\ket{0}^{\otimes{n}} = \ket{\Phi^l}
\end{equation}
where $U_{\Phi^l}$ is a gate that obtains $\ket{\Phi^l}$ from $\ket{0}^{\otimes{n}}$ %=\ket{00..00}$ 
using the Hypergraph States Generation Subroutine~ \cite{quantum_neuron_paper}.
 This moves the quantum state to:
\begin{equation}
\label{eq:state_1}
    \ket{\gamma_1} = \ket{0}\ket{\Phi^l}
\end{equation}

\subsubsection{\textbf{The \texorpdfstring{$\Psi$}{Lg} Entangling Stage}} The second step is to apply another unitary transformation $U_{\Psi}$ to the data register of the circuit in quantum state $\ket{\gamma_1}$. The quantum state moves to:

\begin{equation}
\label{eq:state_2}
    \ket{\gamma_2} = \ket{0}U_{\Psi}\ket{\Phi^l} = \ket{0}\ket{\eta}
\end{equation}
where $\ket{\eta} = U_{\Psi}\ket{\Phi^l}$.

The $U_{\Psi}$ is a gate that moves a quantum state $\ket{\Psi}$ to state $\ket{1}^{\otimes{n}}$, i.e.: 
\begin{equation}
\label{eq:oracle_w}
    U_{\Psi}\ket{\Psi} = \ket{1}^{\otimes{n}} = \ket{N-1}
\end{equation}

Similar to $U_{\Phi}$, the $U_{\Psi}$ gate can be realized using the Hypergraph States Generation Subroutine~ \cite{quantum_neuron_paper}.

\subsubsection{\textbf{The Measurement Stage}} The last step is to measure the neuron activation value, i.e. the dot product $\braket{\Psi|\Phi^l}$. Using equations \ref{eq:state_2} and \ref{eq:oracle_w}, the required similarity can be written as:

\begin{equation}
\label{eq:neuron_activation_with_unitary}
    \braket{\Psi|\Phi^l} = \braket{\Psi|U_{\Psi}^{\dag}U_{\Psi}|\Phi^l} = \braket{N-1|\eta}
\end{equation}

Since %$\bra{N-1}$ is the basis vector $\bra{1}^{\otimes{n}}=\bra{11..11}=[0, 0, ..., 1]$, and 
$\ket{\eta}=\eta_0\ket{0} + \eta_1\ket{1} + ... + \eta_{N-1}\ket{N-1}$,  Equation~\ref{eq:neuron_activation_with_unitary} reduces to

\begin{equation}
\label{eq:neuron_activation_with_unitary2}
    \braket{\Psi|\Phi^l} = \eta_{N-1}
\end{equation}

Hence, the squared neuron activation value is the probability of measuring the data register in state $\ket{N-1}$. %=\ket{1}^{\otimes{n}}$. 
 To measure this probability, we add the final \textrm{CNOT} gate that flips the ancilla qubit if the data register is in state $\ket{N-1}=\ket{1}^{\otimes{n}} =[0, 0, ..., 1]$,

\begin{equation}
\label{eq:final_equaiton}
    |\braket{\Psi|\Phi^l}|^2 = P(a=1) = \eta_{N-1}^2
\end{equation}

%%%%%%%%%%%%%%%%%%%%%%%%%%%%%%%%
\subsection{Example}
In this section, we give a simple example of the quantum binary neuron circuit mentioned in the previous section using two fingerprint vectors collected at two locations: $l_0$ and $l_1$, and a user sample vector, each with four signal sources: $S_i$ for $i \in 
 \{0 .. 3\}$. \\
%Assuming that we have two fingerprint locations $l_0$ and $l_1$, and four signal sources: $S_i$ for $i \in [0 .. 3]$. 
At location $l_0$, all signal sources are heard except $S_3$, i.e. the fingerprint sample at location $l_0$ is: $\Phi^{l_0}=[1,1,1,-1]$, and at location $l_1$, signal sources $S_2$ and $S_3$ are not heard, i.e. fingerprint sample at location $l_1$ is $\Phi^{l_1}=[1,1,-1,-1]$, where $1$ means that the signal source is being heard at the location and $-1$ means that it is not heard. Let the user sample be $\Psi=[1,1,1,1]$, i.e. all signal sources are heard at the current online user location. 

%%%%%%%%%%%%%%%%%%%%%%%%%%%%%%%%%%%%%Example Circuit%%%%%%%%%%%%%%%%%%%%%%%%%%%%%%
\begin{figure}[htp]

\subfloat[Quantum circuit for calculating activation value for the first fingerprint sample $\Phi^{l_0}={[1,1,1,-1]}$ and the user sample $\Psi={[1,1,1,1]}$]{%
    \centering
    \begin{adjustbox}{width=0.4\textwidth}
        \begin{tikzpicture}[row sep=0.1cm]
			\node[text width=1.5cm] at  (-4.8,  1.35){Ancilla};
			\node[text width=1.5cm,rotate=90] at  (-5.0,  0.25){\large Data};
                \node[scale=1.0] {
        \begin{quantikz}
            \lstick{$\ket{0}$}  & \qw       & \qw    & \qw      & \qw      & \targ{}   & \qw &\meter{}& \cw \rstick{$P(a_{0})$}\\
            \lstick{$\ket{0}$}   & \gate{H} \gategroup[wires=2, steps=2, style={dotted, cap=round, inner sep=0pt}, label style={label position=below, yshift=-0.5cm}]{$U_{\Phi^{l_0}}$} & \ctrl{1}  & \gate{H} \gategroup[wires=2, steps=2, style={dotted, cap=round, inner sep=0pt}, label style={label position=below, yshift=-0.5cm}]{$U_{\Psi}$} & \gate{X} & \ctrl{-1} & \qw & \qw & \qw\\
            \lstick{$ \ket{0}$}  & \gate{H} & \gate{Z} & \gate{H} & \gate{X} & \ctrl{-1} & \qw & \qw & \qw
        \end{quantikz}};
        \end{tikzpicture}
    \end{adjustbox}
}

\subfloat[Quantum circuit for calculating activation value for the second fingerprint sample $\Phi^{l_1}={[1,1,-1,-1]}$ and the user sample $\Psi={[1,1,1,1]}$]{%
    \centering
    \begin{adjustbox}{width=0.4\textwidth}
    \begin{tikzpicture}[row sep=0.1cm]
			\node[text width=1.5cm] at  (-4.8,  1.35){Ancilla};
			\node[text width=1.5cm,rotate=90] at  (-5.0,  0.25){\large Data};
                \node[scale=1.0] {
        \begin{quantikz}
            \lstick{$\ket{0}$}   & \qw       & \qw   & \qw      & \qw  & \targ{}   & \qw &\meter{}& \cw \rstick{$P(a_{1})$}\\
            \lstick{$\ket{0}$}   & \gate{H} \gategroup[wires=2, steps=2, style={dotted, cap=round, inner sep=0pt}, label style={label position=below, yshift=-0.5cm}]{$U_{\Phi^{l_1}}$} & \qw  & \gate{H} \gategroup[wires=2, steps=2, style={dotted, cap=round, inner sep=0pt}, label style={label position=below, yshift=-0.5cm}]{$U_{\Psi}$} & \gate{X} & \ctrl{-1} & \qw & \qw & \qw\\
            \lstick{$ \ket{0}$}  & \gate{H} & \gate{Z} & \gate{H} & \gate{X} & \ctrl{-1} & \qw & \qw & \qw
        \end{quantikz}};
        \end{tikzpicture}
    \end{adjustbox}
}

\caption{Binary neuron quantum circuits to calculate neuron activation values between the fingerprint samples $\Phi^{l_i}$ and the user sample $\Psi=[1,1,1,1]$.}
\label{fig:example}
\end{figure}
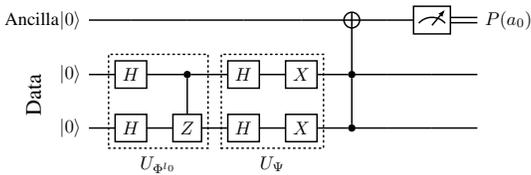
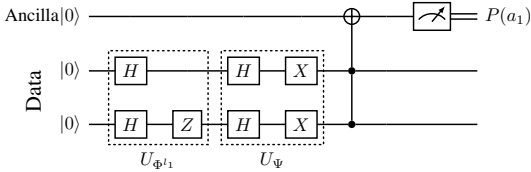

To find the neuron activation value for the fingerprint samples and the user sample vectors, we construct the two quantum circuits shown in Figures~\ref{fig:example}. The first subfigure (a) shows the simple implementation of the $U_{\Phi^{l_0}}$ and $U_{\Psi}$ to find the neuron activation value (dot product) of the first fingerprint sample $\Phi^{l_0}$ with the user sample $\Psi$. The $U_{\Phi^{l_0}}$ gate is implemented by applying the Hadamard gates (H) on the data register, which produces an equal superposition state of $\frac{1}{2}(\ket{00} + \ket{01}+\ket{10} + \ket{11})$. Then the Controlled Z (CZ) gate is added to flip the sign of the basis state $\ket{11}$ by rotating the qubits around the Z-axis of the Bloch sphere~\cite{qbible}. 
 The $U_\Psi$ gate moves state $\ket{\Psi}$ to state $\ket{1}^{\otimes{n}}$ according to Equation~\ref{eq:oracle_w}. This is done by applying H gates to move the $\ket{\Psi}=\frac{1}{2}(\ket{00} + \ket{01}+\ket{10} + \ket{11})$ state to state $\ket{0}^{\otimes{n}}$ then adding NOT gates (X) to move it to state $\ket{1}^{\otimes{n}}$.

The second subfigure (b) shows the quantum circuit implementation for calculating the similarity to the second fingerprint $\Phi^{l_1}$ sample. Similar to the first circuit, the $U_{\Phi^{l_1}}$ is implemented by applying H gates on the data register qubits then a Z gate on the second data qubit to move the data register to state $\frac{1}{2}(\ket{00} + \ket{01}-\ket{10} - \ket{11})$. 

To measure the probability $P(a_0=1)$, we run the first quantum circuit for $k$ times (number of shots) and count the number of times where the ancilla qubit output is  $1$. Then using Equation~\ref{eq:count_equation}, we can find the neuron activation value.

\begin{equation}
\label{eq:count_equation}
    |\braket{\Psi|\Phi^l}|^2 = P(a=1) = \frac{count(a=1)}{k} \simeq 0.5067
\end{equation}

Similarly, we run the second circuit (b) for the same number of shots ($k$) to get the activation value for the location $l_1$ as $\braket{\Psi|\Phi^{l_1}} = 0 $. Since the activation value of the first fingerprint $\braket{\Psi|\Phi^{l_0}}$ is higher than the activation value of the second fingerprint $\braket{\Psi|\Phi^{l_1}}$, we estimate the user's floor as the floor number of the first fingerprint sample's location $l_0$.

\subsection{Complexity Analysis}

In the offline phase, to store a sample vector of size $N$, we need $N$ memory units in a classical computer. This makes the total space complexity needed to store the fingerprint data with $M$ samples $o(MN)$ in the classical algorithm. In contrast, the quantum floor localization algorithm stores a vector sample of size $N$ in $q= 1 + \log(N)$, where one qubit is used as an ancilla and $\log(N)$ qubits register is used as the data register, i.e. the quantum algorithm has a space complexity of order $o(M\log(N))$.

In the online phase, the fingerprint and user vectors are encoded and entangled in the quantum circuit using $U_{\Phi^l}$ and $U_{\Psi}$ unitary transformations, which can be achieved efficiently using the Hypergraph States Generation Subroutine~\cite{quantum_neuron_paper}. Hence, using efficient unitary transformations, the proposed quantum algorithm achieves $o(M\log(N))$ time bound.\\
Furthermore, the fingerprint data vector $\Phi^l$ can be loaded directly from quantum sensors~\cite{qsensor4} in the future or from a quantum random access memory (QRAM) \cite{qram1}, where the binary representation of $\Phi^l_i$ is loaded in parallel into the quantum data register, which allows us to use a single quantum circuit for matching all fingerprint samples with a user's sample.

%%%%%%%%%%%%%%%%%%%%%%%%%%%%%%%%%%%%%%%%%%%%%%%%%%%%%%%%%%%%%%%%%%%%%%%%%%%%%%%%%%%%%%%%%%%%%%%%%%%%%%%%%%%%%%%%%%%%%

\begin{figure*}[!t]
	\centerline
	{\includegraphics[width=0.9\textwidth]{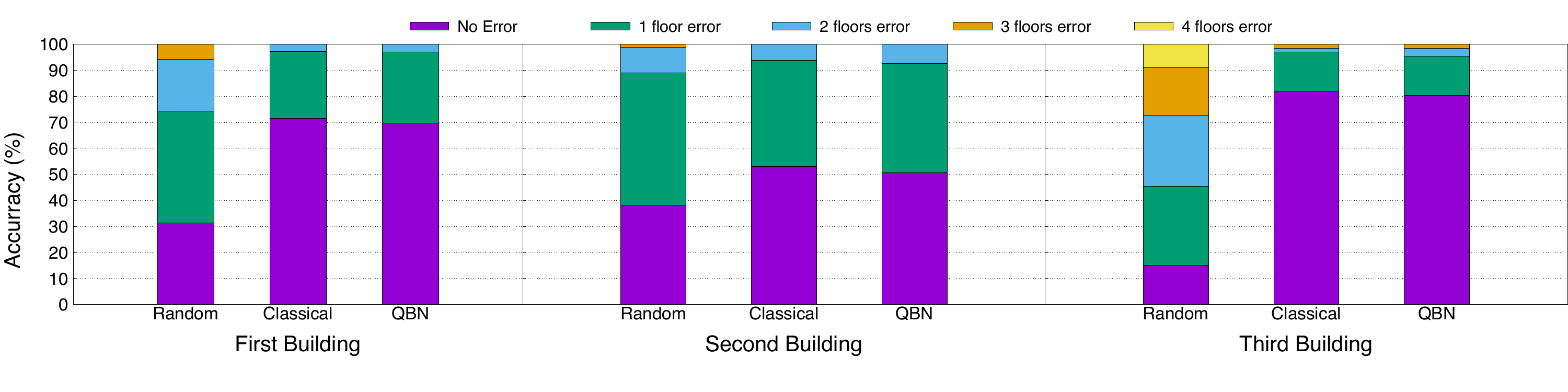}}
	\caption{Floor localization error CDF comparing the proposed quantum  algorithm, the classical counterpart, and a random classifier (as a baseline) for the  three testbeds.}
	\label{fig:b_total}
\end{figure*}

\begin{figure}[!t]
	\centerline
	{\includegraphics[width=0.4\textwidth]{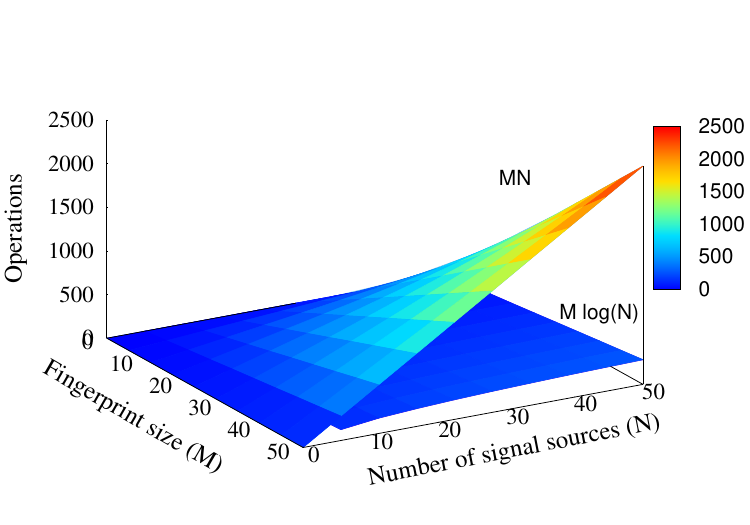}}
	\caption{The complexity of the classical algorithm ($o(MN)$) and the proposed quantum algorithm ($o(MLog(N))$).}
	\label{fig:complexity}
\end{figure}

\begin{figure*}[!t]
	\centerline
	{\includegraphics[width=0.9\textwidth]{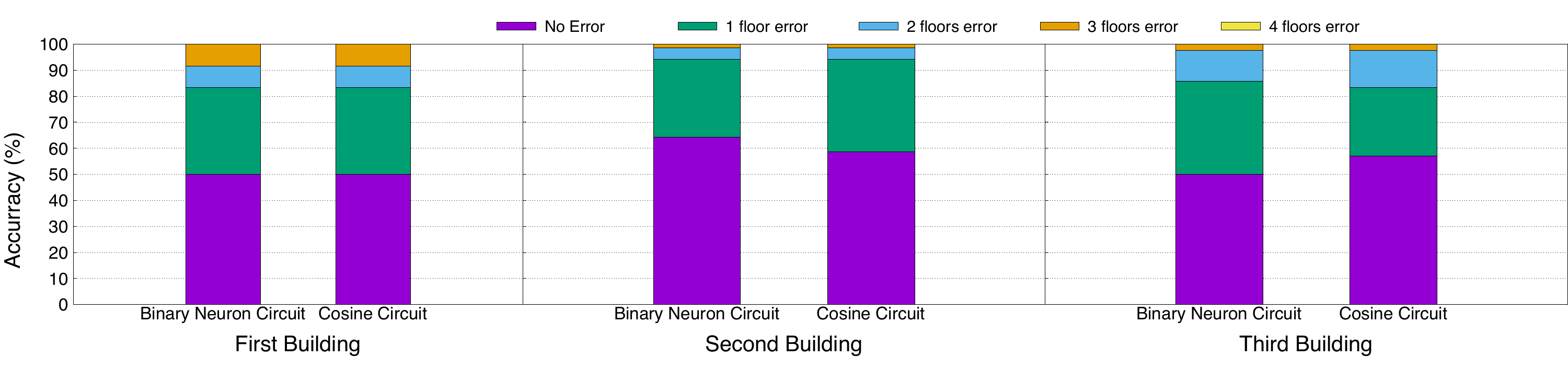}}
	\caption{Floor localization error CDF comparing using the quantum binary neuron circuit, and the quantum circuit mentioned in \cite{quantum_lcn, quantum_qce, quantum_arx} which requires nearly double the number of qubits compared to the proposed quantum binary neuron circuit.}
	\label{fig:cosine_compare}
\end{figure*}

\begin{figure}[!t]
	\centerline
	{\includegraphics[width=0.4\textwidth]{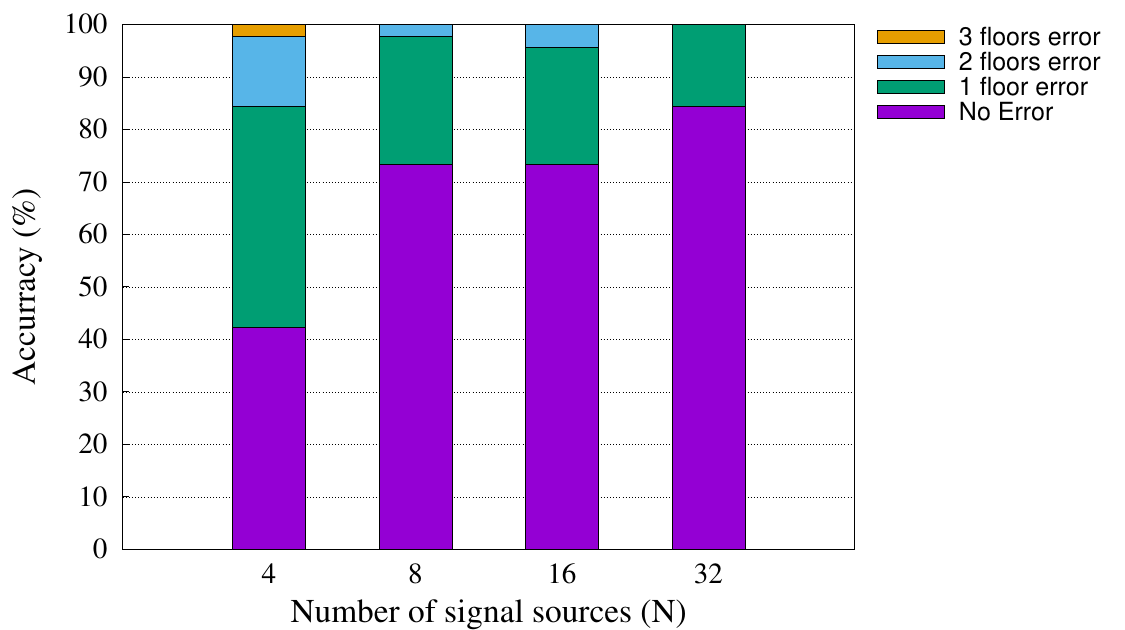}}
	\caption{Floor localization accuracy using different number of signal sources ($N$).}
	\label{fig:diff_n}
\end{figure}

%%%%%%%%%%%%%%%%%%%%%%%%%%%%%%%%%%%%%%%%%%%%%%%%%%%%%%%%%%%%%%%%%%%%%%%%
\section{Evaluation}
\label{sec:evaluation}

In this section, we evaluate the quantum binary neuron algorithm in three real testbeds and implement it on a real IBM quantum machine. We start by describing the testbeds. Then, we show the results of comparing the proposed quantum algorithm with its classical counterpart as well as the state-of-the-art quantum fingerprint localization systems ~\cite{quantum_lcn, quantum_qce, quantum_arx}. Finally, we discuss the different advantages of the proposed quantum algorithm.

\subsection{Environment Setup}
 The collected data cover three different buildings of Jaume~I University~\cite{data_paper}. We deploy the proposed quantum algorithm on the first building (four floors), the second building (four floors), and the third building (five floors). We used the already deployed 16 WiFi access points (APs) as signal sources ($N=16$). The floors area is around $110\textrm{m}^{2}$~\cite{data_paper}.

 We deployed the proposed quantum algorithm on each building and compared the results to the classical counterpart. Furthermore, we compared the proposed quantum circuit to the state-of-the-art quantum fingerprint localization circuit \cite{quantum_lcn,quantum_qce}. We used the real 5-qubits \textit{ibmq\_manila} quantum computer~\cite{ibmq}.

\subsection{Performance Evaluation}
\subsubsection{\textbf{Comparison to classical binary neuron}} Figure~\ref{fig:b_total} shows the floor localization accuracy CDF for the three buildings using the Quantum Binary Neuron (QBN) algorithm, the classical counterpart, and the random floor classifier as a worst-case bound for localization. The figure shows that both the quantum and classical algorithms achieve the same accuracy. This comes with an exponential gain in both the running time and the space.

Figure~\ref{fig:complexity}, shows the space and running time complexity between the classical and the proposed quantum algorithm. The figure shows that the proposed quantum algorithm achieves $o(M\log(N))$ compared to $o(MN)$. This highlights the exponential gain of the proposed quantum algorithm over its classical counterpart due to the fact that the quantum algorithm grows in a logarithmic manner with $N$.

\subsubsection{\textbf{Comparison to state-of-the-art quantum localization}} To validate that the quantum binary neuron gives similar accuracy to the state-of-the-art quantum localization~\cite{quantum_lcn, quantum_qce, quantum_arx} while using a lower number of qubits, we run the quantum binary neuron circuit and the  circuit used in the state-of-the-art quantum algorithm ~\cite{quantum_lcn, quantum_qce, quantum_arx} on the same testbeds. However, we use only \textbf{four APs}  in this experiment, as the previous  quantum algorithms would require more than the 5-qubits available on the used \textit{ibmq\_manila} quantum machine. This highlights the advantage of our proposed algorithm that can use up to $16$ APs on this machine leading to better localization accuracy and making it more suitable for near-term quantum devices. 

Figure~\ref{fig:cosine_compare} validates that the quantum binary neuron circuit can achieve a similar accuracy to the quantum circuit used in the state-of-the-art quantum localization  algorithms~\cite{quantum_lcn, quantum_qce, quantum_arx}. The proposed quantum binary neuron requires almost half the number of qubits ($1+\log(N)$) compared to the previous quantum algorithms that require ($1+2\log(N)$) qubits.

Moreover, Figure~\ref{fig:diff_n}, confirms that the higher the number of signal sources, the more precise the accuracy will be. This highlights the importance of using more signal sources, which can be achieved in low number of qubits ($1+\log(N)$) using the proposed quantum binary neuron algorithm. This makes it more feasible to be implemented on near-term quantum devices.

\subsection{Discussion}
The results of running our quantum algorithm on three real testbeds show that the proposed quantum floor localization algorithm has a similar accuracy as its classical counterpart but with the exponential gain in time and space, and has a similar accuracy to the state-of-the-art quantum algorithms~\cite{quantum_lcn, quantum_qce, quantum_arx} while using a lower number of qubits.

The classical floor estimation algorithm requires $o(MN)$ space and running time. On the other side, the proposed quantum algorithm requires $o(M\log(N))$ space and time. This highlights the quantum computing gain and enables fingerprint-based floor localization systems to be deployed in large-scale multi-story buildings while maintaining a similar localization accuracy as the classical algorithm \cite{bahl2000radar, youssef2005horus}. 

Furthermore, the proposed quantum algorithm is useful for the offline calibration phase as it reduces the fingerprint vector size from $o(N)$ to $o(\log(N))$, which mitigates the data storage overhead and saves the required bandwidth for applications where the data is downloaded on the users' devices when the matching process is done on users' local devices (e.g., for privacy-sensitive applications). 

Comparing the proposed quantum algorithm to the state-of-the-art quantum localization algorithms that require $1+2\log(N)$ number of qubits, results show that the proposed quantum binary neuron algorithm achieves similar accuracy while using $1+\log(N)$ qubits only compared to these algorithms. This makes it able to achieve higher accuracy on near-term quantum machines, where the number of qubits is limited. %, compared to the state-of-the-art quantum algorithms.

Finally, a single quantum circuit can be used for different fingerprint samples by loading the fingerprint data from a quantum random access memory (QRAM) or quantum sensors instead of the unitary transformation $U_{\Phi^l}$ in the first stage, and both space and time complexity can be further enhanced to $o(\log(MN))$ by encoding all the fingerprint locations using quantum states. And state preparation routines could be used to reduce the number and depth of the circuit gates. 

%%%%%%%%%%%%%%%%%%%%%%%%%%%%%%%%%%%%%%%%%%%%%%%%%%%%%%%%%%%%%%%%%%%%%%%%%%%%%%%%%%%%%%%%%%%%%%%%%%%%%%%%%%%%%%%%%%%%%
\section{Conclusion}
\label{sec:conclusion}
 In this paper, we introduced a fingerprint-based quantum algorithm for multi-story large-scale floor localization. 
The proposed quantum algorithm requires $o(M\log(N))$ time and space, unlike its classical counterpart, which needs $o(MN)$ time and space for a fingerprint with $M$ locations and $N$ signal sources. 
 We implemented the proposed quantum algorithm on a real IBM quantum machine and deployed it in three real testbeds. Results show that our quantum algorithm can achieve the same accuracy as the classical algorithm but with the exponential gain in time and space. This confirms the gain of using quantum algorithms to solve large-scale floor localization problems, where more data can be used with lower number of computing resources. Furthermore, we show the advantage of using the proposed quantum binary neuron circuit for near-term quantum machines where the number of qubits is limited. The proposed algorithm can use nearly half the number of qubits compared to other state-of-the-art quantum localization algorithms while keeping the localization accuracy.
Currently, we are expanding our research in multiple directions including exploring different quantum similarity metrics implementations, extending our exponential saving to the fingerprint samples ($M$), handling the intrinsic practical considerations of quantum processors in the algorithm development, obtaining theoretical quantum performance bounds, among others.

\balance
\bibliography{main}
\end{document}